\documentclass[a4paper,fleqn,usenatbib]{mnras}
\usepackage{ulem}
\newcommand{\arxiv}{arXiv:[arXiv:1107.5325]}

\newcommand{\Zeitschrift}{Z. Nat. Teil A}
\newcommand{\Living}{Sol. Phys}

\usepackage[T1]{fontenc}
\usepackage{ae,aecompl}

\usepackage{graphicx}	
\usepackage{amsmath}	
\usepackage{amssymb}	

\usepackage{color}
\usepackage{ulem}

\title[$\zeta^{1} + \zeta^{2}$ Ret: A puzzling activity pattern]
{$\zeta^{1} + \zeta^{2}$ Reticuli binary system: A puzzling chromospheric activity pattern\thanks{Based on
data products from observations made with ESO Telescopes at the La Silla Paranal Observatory.}}

\author[M. Flores et al.]{
M. Flores,$^{1,3,6}$\thanks{E-mail: matiasflorestrivigno@conicet.gov.ar}
C. Saffe,$^{1,3,6}$
A. Buccino,$^{2,4,6}$
M. Jaque Arancibia,$^{1,6,7}$
J. F. Gonz\'alez,$^{1,3,6}$ 
\newauthor
N. E. Nu\~nez$^{1,3,6}$
and  E. Jofr\'e$^{5,6}$\\
$^{1}$Instituto de Ciencias Astron\'omicas, de la Tierra y del Espacio (ICATE), Espa\~na Sur 1512,
    CC 49, 5400 San Juan, Argentina.\\
$^{2}$Instituto de Astronom\'ia y F\'isica del Espacio (IAFE), Buenos Aires, Argentina.\\
$^{3}$Facultad de Ciencias Exactas, F\'isicas y Naturales, Universidad Nacional de San Juan, San Juan, Argentina.\\
$^{4}$Departamento de F\'isica, Facultad de Ciencias Exactas y Naturales, Universidad de Buenos Aires,
   Buenos Aires, Argentina.\\
$^{5}$Observatorio Astron\'omico de C\'ordoba (OAC), Laprida 854, X5000BGR, C\'ordoba, Argentina\\
$^{6}$Consejo Nacional de Investigaciones Cient\'ificas y T\'ecnicas (CONICET), Argentina.\\
$^{7}$Departamento de F\'isica y Astronom\'ia, Universidad de La Serena, Av. Cisternas 1200, La Serena, Chile
}

\date{Accepted XXX. Received YYY; in original form ZZZ}

\pubyear{2017}

\begin{document}
\label{firstpage}
\pagerange{\pageref{firstpage}--\pageref{lastpage}}
\maketitle

\begin{abstract}
We perform, for the first time, a detailed long-term activity study of the binary system $\zeta$ Ret. We use all
available HARPS spectra obtained between the years 2003 and 2016. We built a time series of the Mount Wilson $S$ index for both stars, then we analyse these series by using Lomb-Scargle periodograms. The components $\zeta^{1}$ Ret and $\zeta^{2}$ Ret that belong to this binary system are physically very similar to each other and also similar to our Sun, which makes it a remarkable system.  We detect in the solar-analogue star $\zeta^{2}$ Ret a long-term activity cycle with a period of $\sim$10 yr, similar to the solar one ($\sim$11 yr). It is worthwhile to mention that this object satisfies previous criteria for a flat star and for a cycling star simultaneously. Another interesting feature of this binary system, is a high $\sim$0.220 dex difference between the averages
log($\mathrm{R}'_\mathrm{HK}$) activity levels of both stars. Our study clearly shows that $\zeta^{1}$ Ret is significantly more
active than $\zeta^{2}$ Ret. In addition, $\zeta^{1}$ Ret shows an erratic variability in its stellar activity. In this work, we explore different scenarios trying to explain this rare behaviour in a pair of coeval stars, which could help to explain the difference in this and other binary systems. From these results, we also warn that for the development of activity-age calibrations (which commonly use binary systems and/or stellar clusters as calibrators) it should be taken into account the whole history of activity available of the stars involved. 
\end{abstract}



\section{Introduction}
Stellar activity studies have several applications. They can be used to disentangle stellar and planetary signals
in radial-velocity surveys \citep[e.g.][]{2013ApJ...774..147R,2014A&A...567A..48C,2016A&A...585A.134D}. Also, studying the long-term  magnetic activity  of stars with  physical characteristics similar to the Sun (T$_\mathrm{eff}$, log g, age, etc.) could help to place our Sun in context \citep[e.g.][]{2007AJ....133..862H,2009AJ....138..312H}.
 In addition, they can be used to better understand the star-planet interactions
 \citep[e.g.][]{2011A&A...530A..73C,2012A&A...540A..82K,2015ApJ...799..163M}.

First stellar activity studies were initiated by Olin Wilson in 1966, which gave rise to the HK project
 \citep[][]{1978ApJ...226..379W}. This program was carried out at Mt. Wilson Observatory and continued until 2003. It has been the main source of many activity-related projects  \citep[e.g.][]{1978PASP...90..267V,1991ApJS...76..383D,1995ApJ...441..436G,1998ASPC..154..153B}. These works allowed better understanding of the stellar magnetic phenomena beyond the solar activity. By using the \ion{Ca}{ii} H\&K line cores as activity proxies through the standard Mount Wilson index ($S$), \citet{1998ASPC..154..153B} grouped the sample of 2200 stars  in three classes corresponding to different long-term activity behaviour. They found that stars with moderate activity showed cycles with periods between 2.5 and 21 yr, while very active stars displayed fluctuations of activity rather than cycles, in general, corresponding to young stars with high rotation velocities. Finally, inactive stars are found in a phase that may be similar to the solar Maunder Minimum\footnote{Period between 1645 and 1715 where solar activity was greatly reduced \citep[e.g.][]{1976Sci...192.1189E}.} (hereafter MM).

Currently, new activity cycles have been reported by several authors
\citep[e.g.][]{2010ApJ...723L.213M,2010ApJ...722..343D,2014ApJ...781L...9B,2015ApJ...812...12E}.  
These activity cycles have been found in main sequence and even post main sequence stars with spectral types ranging from F to M \citep{1998ASPC..154..153B}, including solar analogue/twins and stars with planets \citep[e.g.][]{2016A&A...589A.135F}. In addition, several stars seem to have multiple activity cycles \citep[e.g.][]{1995ApJ...438..269B,2007AJ....133..862H,2009A&A...501..703O,2013ApJ...763L..26M,2015ApJ...812...12E,2017MNRAS.464.4299F}.

Samples of binary systems were included into stellar activity studies and also used to obtain activity-rotation-age calibrations. 
Observational data \citep{2007AJ....134.2272R,2012A&A...546A..63V,2012MNRAS.421.2025K} and numerical simulations \citep[e.g.][]{2012Natur.492..221R} support the idea that most of binary stars are formed from a common molecular cloud (i.e., coeval stars), therefore it is expected that components of binary systems present similar properties such as age and chemical composition.
As stellar activity is thought to be produced by a global-scale dynamo action which is powered by the rotational velocity (or angular momentum) and turbulent convection 
\citep[e.g.][]{1955ApJ...122..293P,1966ZNatA..21..369S,1982AA...108..322,2017Sci...357..185S}, it decays during stellar life time as a consequence of magnetic braking and angular momentum loss due to stellar wind. Then, a difference in activity between similar components of a binary system led to some initial interpretations such as the measurement of different phases of long-term variations \citep[e.g.][]{1995ApJ...438..269B,1998ASPC..154..153B,1998ASPC..154.1235D},
one component possibly leaving the main sequence, or the rotational modulation \citep{1998ASPC..154.1235D}, nowadays the origin of the difference in activity is not fully understood.

By using a sample of binary systems and star cluster members, \citet[][]{2008ApJ...687.1264M} (hereafter MH08) derived an improved activity-age relation which allows to obtain a chromospheric age by means of the chromospheric activity measured from \ion{Ca}{ii} H\&K emission lines. To do so, they also took into account the relation between $\mathrm{R}'_\mathrm{HK}$ and colour (i.e., mass), 
which until that time had not been considered \citep[e.g.][]{1991ApJ...375..722S,1993PhDT.........3D,1999A&A...348..897L}. However, the validity of these activity-age calibrations is still under discussion in the literature \citep[e.g.][]{2013A&A...551L...8P,2016A&A...594L...3L}.

In 2015, we started a program aiming to study stellar activity in solar-analogue and solar-twin stars using the extensive database of HARPS spectra. 
Our initial sample comprised solar-twin stars taken from \citet{2015A&A...579A..52N}. Currently, we have extended our sample to other particular stars and also to binary systems with similar components. Some results of this young program have been recently published 
\citep[see][for details]{2016A&A...589A.135F,2017MNRAS.464.4299F}.     

One remarkable object in our current sample is the $\zeta$ Reticuli binary system (hereafter $\zeta$ Ret).
A Bayesian analysis of the proper motions indicates a very high probability (near 100\%) that the pair is physically connected \citep{2011ApJS..192....2S}. The spectral types of their components are very similar with each other (G2 V + G1 V as referred in the Hipparcos database) and also similar to the Sun, being both solar analogues \citep[see][for details]{2016A&A...588A..81S}. In addition, the (B$-$V) colours of the stars $\zeta^{1}$ Ret and $\zeta^{2}$ Ret are 0.64 $\pm$ 0.01 and 0.60 $\pm$ 0.01 mag, according to \citet{1997A&A...323L..49P}. This makes $\zeta$ Ret a unique laboratory that allows to carry out a detailed test of stellar activity. $\zeta$ Ret is a wide binary system, with a separation of $\sim$3700 AU \citep{2001AJ....122.3466M}, which discard any physical interaction between the components. The star $\zeta^{2}$ Ret (= HD 20807) has a debris disc at $\approx$ 100 AU, which was detected through a mid-IR excess \citep{2008ApJ...674.1086T} and then confirmed by direct imaging \citep{2010A&A...518L.131E}. In contrast, Spitzer and Herschel
observations of $\zeta^{1}$ Ret (= HD 20766) have not reveal the presence of IR excess \citep{2006ApJ...636.1098B,2008ApJ...674.1086T,2013A&A...555A..11E}.
Additionally, both stars have been monitored by the Anglo-Australian Planet Search (AAPS) radial-velocity
survey\footnote{\url{http://newt.phys.unsw.edu.au/~cgt/planet/Targets.html}}, and included in the ESO CES and HARPS GTO planet search
programs \citep[e.g.][]{2008A&A...487..373S,2013A&A...552A..78Z} giving, at the moment, no-planet
detection.

There is a previous stellar activity study of the star $\zeta^{2}$ Ret in the literature, which was carried out by \citet{2011arXiv1107.5325L}. In this work the authors report 99 stars with magnetic cycles, finding a period of $1133_{-65}^{+1090}$ days for $\zeta^{2}$ Ret by using more than 6 yr of observations ($\zeta^{1}$ Ret was not analyzed). Fortunately, the HARPS monitoring of both components of this binary system continued, providing now a large data set that can be used to make a detailed long-term stellar activity study in this system. In this way, we are starting to study possibly different long-term activity levels observed in binary systems with similar components,
which are difficult to explain. In addition, solar-analogue nature of both  stars $\zeta^{1}$ Ret and $\zeta^{2}$ Ret  could also be used to compare them directly with our Sun,
what has been done for few stars such as 18 Sco \citep{2007AJ....133..862H} and HD 45184 \citep{2016A&A...589A.135F}. Recently, solar activity is under discussion in the stellar context. Given its activity and rotation period,  many authors  suggest that the Sun could be in a transitional phase with an atypical activity cycle in the stellar dynamo theory \citep{2007ApJ...657..486B,2016ApJ...826L...2M}. All these works encouraged the present study of the $\zeta$ Ret binary system.

This study is organized as follows. In Section \S\ref{sec.obs} we describe
the observations and data reduction, while in Section \S\ref{sec.res} we describe our main results.  In Section \S\ref{sec.disc} we provide a discussion, and finally in Section \S\ref{sec.concl} we summarize our main conclusions.

\section{Observations and data reduction}
\label{sec.obs}

Stellar spectra of the binary system $\zeta$ Ret were taken between 
2003 and 2016 with the HARPS (High
Accuracy Radial velocity Planet Searcher) spectrograph attached to the La Silla 3.6m ESO telescope. These public high-resolution spectra (R $\sim$115.000) were obtained from the ESO HARPS archive\footnote{\url{https://www.eso.org/sci/facilities/lasilla/}}, under programs listed in Table \ref{tab.one}. They have been automatically processed by the HARPS pipeline\footnote{\url{http://www.eso.org/sci/facilities/lasilla/instruments/
harps/doc.html}}. These spectra typically cover a spectral range from $\sim$3780 to $\sim$6910 \AA \ and present a signal-to-noise (S/N) $\sim$150 at $\sim$6440 \AA \ for both stars.

\begin{table}
  \centering
  \caption{HARPS ID observing programs used in this work.}
  \label{tab.one}
  \begin{tabular}{@{}cc@{}}
\hline
\hline
\multicolumn{2}{|c|}{ESO HARPS programs}\\
\hline
078.C-0833(A) &  074.C-0364(A) \\                        
077.C-0530(A) &  060.A-9036(A)\\
078.C-0044(A) &  076.C-0878(A)\\
079.C-0681(A) &  074.C-0012(B)\\
072.C-0513(B) &  072.C-0488(E)\\
074.C-0012(A) &  183.C-0972(A)\\
073.C-0784(B) &  074.C-0012(B)\\
072.C-0513(D) &  192.C-0852(A)\\
 \hline
  \end{tabular}
\end{table}

In order to compute the $S$ index, HARPS spectra were corrected by radial velocity by using standard IRAF\footnote{\textsf{IRAF is distributed by the National Optical Astronomy Observatory, which
    is operated by the Association of Universities for Research in
    Astronomy, Inc. under cooperative agreement with the National
    Science Foundation.}} tasks. Then, we discard the spectra with low S/N ($\lessapprox100$) and measured the \ion{Ca}{ii} H\&K line-core fluxes of 337 HARPS spectra (being 68 for $\zeta^{1}$ Ret and 269 for $\zeta^{2}$ Ret). To do so, for each spectrum we integrated the flux in two windows centred at the cores of the \ion{Ca}{ii} lines, weighted with triangular profiles of 1.09 \AA \ full width at half-maximum (FWHM), and computed the ratio of these fluxes to the mean continuum flux, integrated in two passbands of width 20 \AA \ centred at 3891 and 4001 \AA. Finally, we converted this ratio into  
the $S$, following \citet{2011arXiv1107.5325L}, as in our previous works \citep{2016A&A...589A.135F,2017MNRAS.464.4299F}.

We complement our analysis with X-ray observations sensitive in the 0.2-10.0 keV band.
To do so, we search in the HEASARC\footnote{\url{https://heasarc.gsfc.nasa.gov/docs/archive.html}} database using 
the coordinates of both component of the system. As a result, we found only one observation per star   
in the XMM-Newton Serendipitous Source Catalog\footnote{\url{http://xcatdb.unistra.fr/3xmmdr6}}.
These individual measurements correspond to the dates 2006, August 05 and 2002, May 3 for
$\zeta^{1}$ Ret and $\zeta^{2}$ Ret, respectively. In order to obtain X-rays fluxes (F$_X$), source data were extracted from a circular aperture of fixed radius (28$\prime\prime$) centred on the detection position, while background data were accumulated from a co-centred annular region with inner and outer radii of 60$\prime\prime$  and 180$\prime\prime$, respectively. In this way, we estimated F$_X$ values of (5.11 $\pm$ 0.08) 10$^{-13}$  and (0.25 $\pm$ 0.32) 10$^{-13}$ for $\zeta^{1}$ Ret and $\zeta^{2}$ Ret, respectively.


\section{Results}
\label{sec.res}

\subsection{Chromospheric activity of the $\zeta$ Ret binary system}

\begin{figure}
\centering
\includegraphics[width=\columnwidth]{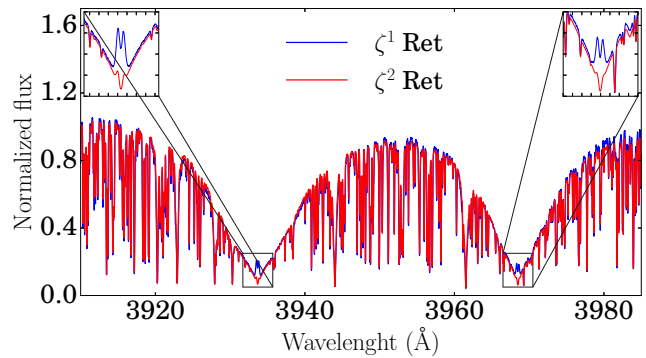}
\caption{Average \ion{Ca}{ii} H\&K profiles for $\zeta^{1}$ Ret (blue solid line) and $\zeta^{2}$ Ret (red solid line). For clarity, we include a zoom in the \ion{Ca}{ii} H\&K spectral cores.}
\label{nar0}
\end{figure}

\begin{figure}
\centering
\includegraphics[width=\columnwidth]{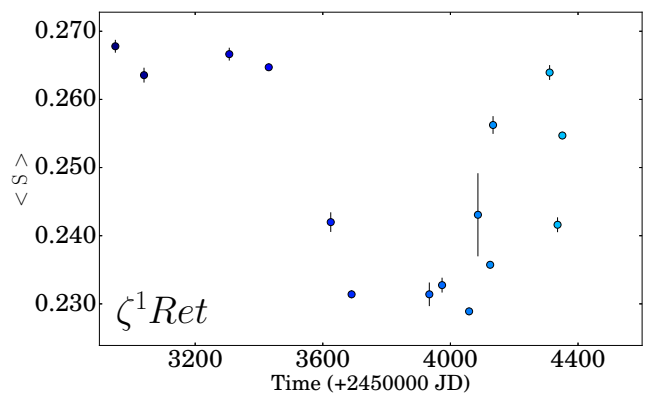}
\caption{Time serie of the Mount Wilson indexes for $\zeta^{1}$ Ret. Different circles colours correspond to the same observing season of Fig. \ref{plotonebinary1}.}
\label{plotonebinary0}
\end{figure}

\begin{figure}
\centering
\includegraphics[width=\columnwidth]{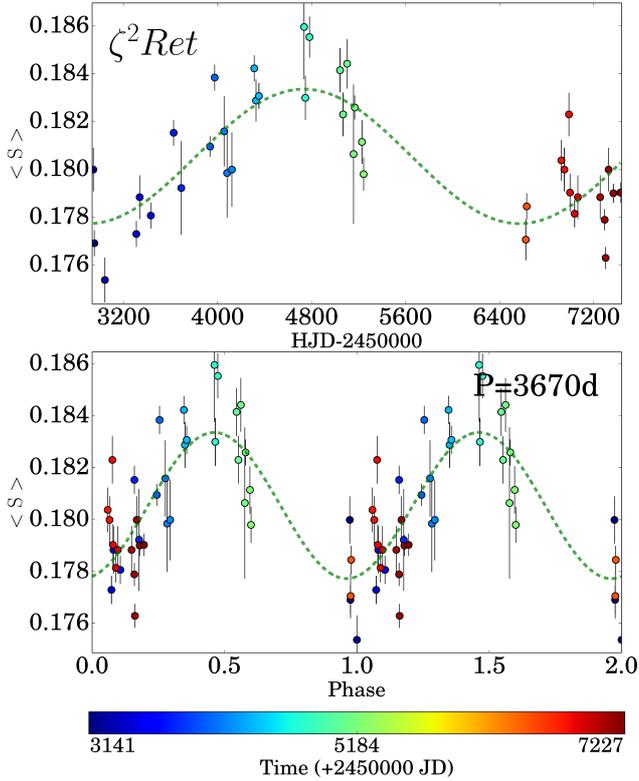}
\caption{Top panel: Time serie of the Mount Wilson index for $\zeta^{2}$ Ret. The dotted green line indicates the cycle calculated in this work. Lower panel: Mount Wilson index phased (showed between 0 and 2) with a 10 yr period.}
\label{plotonebinary1}
\end{figure}

The average of \ion{Ca}{ii} H\&K profiles for both components of $\zeta$ Ret
are shown in Fig.~\ref{nar0}. A clear difference between both stars can be seen from their \ion{Ca}{ii} H\&K line cores. On the other hand, in Figs~\ref{plotonebinary0} and \ref{plotonebinary1} we show the time series for both components. Due to the high sampling frequency of HARPS and in order to diminish the scatter probably produced by rotational modulation of individual active regions \citep[][]{1995ApJ...438..269B}, we have followed the same procedure detailed in our
previous works \citep[][]{2016A&A...589A.135F,2017MNRAS.464.4299F}. Data points are the average values of observations associated to the same observing season (HARPS monthly means), while error bars of the average values correspond to the standard deviation of the mean. We note that $\zeta^{1}$ Ret presents higher activity levels than $\zeta^{2}$ Ret for all (individual or average) measurements.

\begin{figure}
\centering
\includegraphics[width=\columnwidth]{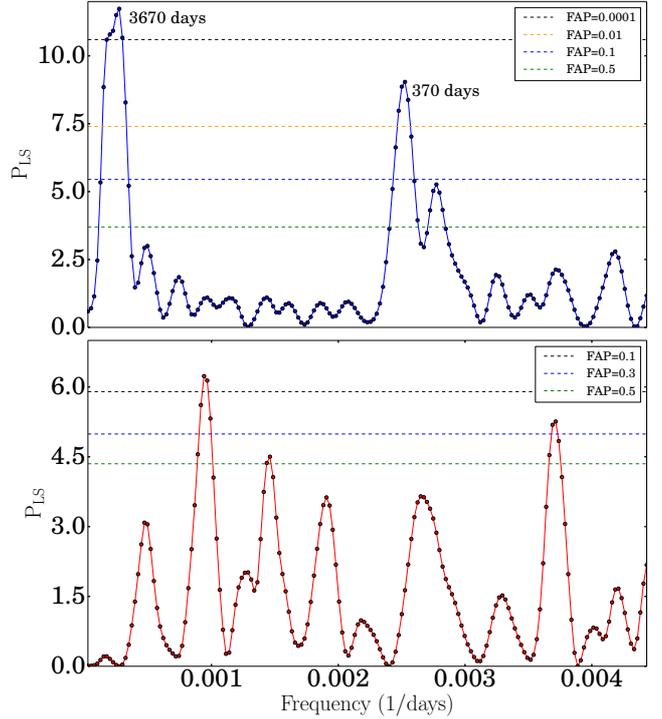}
\caption{Top panel: Lomb-Scargle periodogram of the HARPS data set plotted in Fig.\ref{plotonebinary1}. Lower panel: Lomb-Scargle periodogram after subtracting the  $\sim$10 yr long-term period. Dashed horizontal colour lines correspond to different values of FAP.}
\label{plotonebinary3}
\end{figure}

To study the long-term activity of both components by using the \ion{Ca}{ii} H\&K lines, we compute the Lomb-Scargle periodogram \citep{1986ApJ...302..757H}. As a result, we obtain a long period of 3670 $\pm$ 170 days ($\sim$10 yr) with a Monte Carlo FAP\footnote{See details in \citet{2009A&A...495..287B}} (\textit{False Alarm Probability}) of 3.2$\times10^{-6}$ for $\zeta^{2}$ Ret (see top panel of Fig. \ref{plotonebinary3}.).
We use a cutoff in FAP for the detection of periodicities of 0.1\% (0.001), similar to previous works \citep[e.g.][]{1995ApJ...438..269B,2016A&A...595A..12S}. Together with the most significant peak, appears a short period less significant ($\sim370$ days) that could correspond to an alias. Then this possibility was explored by  subtracting the main periodic variation and recalculating the periodogram, which was carried out following \citet{2016A&A...595A..12S}. The second peak is not present in the new periodogram (Fig.~\ref{plotonebinary3}, lower panel), indicating that it is not a real period. On the other hand, there is no evidence for a long-term activity cycle in $\zeta^{1}$ Ret from a Lomb-Scargle analysis. However, we caution that our temporal coverage for $\zeta^{1}$ Ret is $\sim$3.3 yr. Then, more observations are desirable in order to detect a possible long-term activity cycle, given that up to now this star shows an erratic rather than a cyclic behaviour.

The fitted harmonic curve with a period of $\sim$10 yr calculated in this work was obtained as a least-square fit to the monthly means \citep[see][for details]{2009A&A...495..287B}. It can be seen a good agreement between observational data points and the fitted curve.

From a visual inspection of Figs~\ref{plotonebinary0} and \ref{plotonebinary1}, aclear difference in morphology between both time series is also readily evident. In a previous work, we showed that both stars present very similar stellar parameters \citep[see Table 2 in][]{2016A&A...588A..81S}. In order to compare their chromospheric activity levels, we computed the $\mathrm{R}'_\mathrm{HK}$  index by subtracting the photospheric contribution, following \citet{1984ApJ...279..763N}. The resulting  log($\mathrm{R}'_\mathrm{HK}$) mean values are $-4.64 \pm 0.014$ dex and $-4.86 \pm 0.013$ dex for $\zeta^{1}$ Ret and $\zeta^{2}$ Ret, respectively. The uncertainties in log($\mathrm{R}'_\mathrm{HK}$) were estimated by adding quadratically the error related to (B$-$V) and an additional instrumental noise of $\sim$0.01 dex, which was estimated following \citet{2011arXiv1107.5325L}. Then, according to activity regions defined by \citet{1996AJ....111..439H}, $\zeta^{1}$ Ret corresponds to an active star while $\zeta^{2}$ Ret to an inactive one.

For comparison, in Table \ref{tabone} we show activity measurements from the literature for the $\zeta$ Ret binary system.
It is interesting to note that MH08 used for $\zeta^{2}$ Ret a mean activity index of $-4.79$ dex \citep[taken from][]{1996AJ....111..439H} in order to derive an improved activity-age relation, which is the highest value reported for this star.

\begin{table}
\caption{Activity measurements from the literature for the $\zeta$ Ret binary system.}
\label{tabone}
\begin{tabular}{lcc}
\hline
Reference & $\zeta^{1}$ Ret & $\zeta^{2}$ Ret \\
          &  log($\mathrm{R}'_\mathrm{HK})$                                & log($\mathrm{R}'_\mathrm{HK})$\\
\hline
{\citet{1996AJ....111..439H}}      & -4.65 & -4.79 \\
{\citet{2006MNRAS.372..163J}}    & -4.58 & -4.84 \\
{\citet{2010A&A...520A..79M}}   & -4.86 & $--$  \\
{\citet{2011arXiv1107.5325L}}      & $--$  & -4.90 \\
{\citet{2013A&A...552A..78Z}}   & -4.67 & -4.89 \\ 
{This work} & -4.64 & -4.86 \\  
\hline
\end{tabular}
\end{table}

\begin{figure}
\centering
\includegraphics[width=\columnwidth]{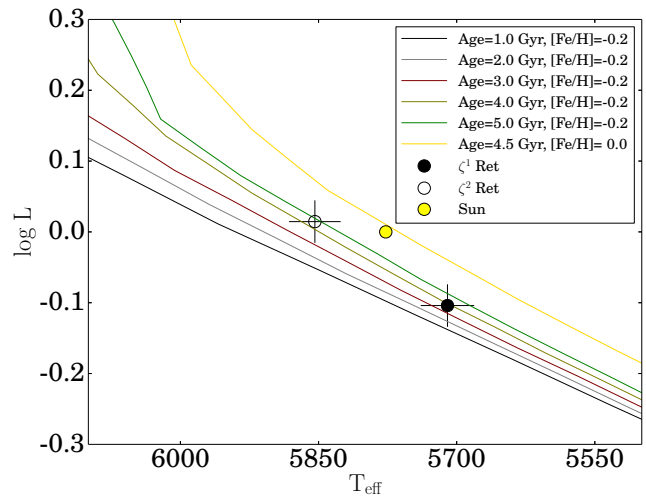}
\caption{Luminosity vs T$_\mathrm{eff}$ for $\zeta^{1}$ Ret (full circle), $\zeta^{2}$ Ret (empty circle) and for the Sun (yellow circle). The continuous lines indicate the isochrones corresponding to different ages and metallicities, as predicted by Yonsei-Yale  \citep{2001ApJS..136..417Y,2004ApJS..155..667D}. For instance, the yellow line correspond to the isochrone with an age of 4.5 Gyr and solar metallicity ([Fe/H]$=$0).}
\label{natyr}
\end{figure}

It would be useful to know the stellar age of this binary system.
To do so, we estimated ages of both stars using Yonsei-Yale isochrones \citep{2001ApJS..136..417Y,2004ApJS..155..667D} as described 
in \citet{2012A&A...543A..29M} and \citet{2013ApJ...764...78R}, employing the q$^{2}$ Python package \citep{2014A&A...572A..48R}.
By adopting the spectroscopic fundamental parameters of \citet{2016A&A...588A..81S}, V magnitudes from the Hipparcos and
Tycho Catalogues \citep{1997ESASP1200.....E}, and revised parallaxes from \citet{2007A&A...474..653V}, we estimated an age of
4.0 $\pm$ 1.7 Gyr and 4.7 $\pm$ 1.4 Gyr for $\zeta^{1}$ Ret and $\zeta^{2}$ Ret, respectively, i.e., an average age 
of $\sim$4.4 Gyr. For comparison, we also derived the ages of both stars using a Bayesian approach through the
PARAM\footnote{\url{http://stev.oapd.inaf.it/cgi-bin/param}} code \citep{2006A&A...458..609D}, using the PARSEC isochrones of
\citet{2012MNRAS.427..127B}. With this method, we find ages of 4.1 $\pm$ 1.9 Gyr and 5.5 $\pm$ 1.1 Gyr for $\zeta^{1}$ Ret
and $\zeta^{2}$ Ret, respectively, in agreement with the previous determination. In the Figure \ref{natyr} we present Yonsei-Yale isochrones  \citep{2001ApJS..136..417Y,2004ApJS..155..667D} corresponding to different ages and metallicities (continuous lines), together with the position of the Sun (yellow circle), the star $\zeta^{1}$ Ret (full circle) and $\zeta^{2}$ Ret (empty circle). The rotational velocities of both stars are $v$ $\sin$ $i$ $\sim$2.7 $\pm$ 0.1 km s$^{-1}$ for $\zeta^{1}$ Ret and $v$ $\sin$ $i$ $\sim$2.7 $\pm$ 0.3 km s$^{-1}$ for $\zeta^{2}$ Ret according to \citet{2003A&A...398..647R}, who caution that very low projected rotational velocities ($v$ $\sin$ $i$ < 3 km s$^{-1}$) must be interpreted as upper limits. These low velocities    
together with the non-detection of the Lithium line at 6707.8 \AA, point toward an age similar to our Sun, while the high activity (particularly of $\zeta^{1}$ Ret) would indicate a younger age for the system.

It is interesting to note that both set of isochrones seems to indicate that $\zeta^{2}$ Ret is slightly older than $\zeta^{1}$ Ret (but they agree within the uncertainties). Then, we explore the possibility that both stars do present different ages. To do so, we estimated the expected activity using the age-activity relations of MH08 in order to compare to the activity measured. Using the ages from Yonsei-Yale, the resulting log($\mathrm{R}'_\mathrm{HK}$) values are -4.95 $\pm$ 0.19 dex and -5.12 $\pm$ 0.15 dex, while using the ages from the PARSEC isochrones we obtained -4.96 $\pm$ 0.21 dex and -5.20 $\pm$ 0.12 dex, for the stars $\zeta^{1}$ Ret and $\zeta^{2}$ Ret, respectively. Then, the expected activity values resulted lower than those observed in both stars ($-4.64 \pm 0.014$ dex and $-4.86 \pm 0.013$ dex for $\zeta^{1}$ Ret and $\zeta^{2}$ Ret). If these expected values were in better agreement with the observed ones, this would give some support to the idea that $\zeta$ Ret is not a coeval system.

\section{Discussion}
\label{sec.disc}

\subsection{Possible scenarios to explain the activity differences in binary systems}
Up to now, a difference in the activity level between similar stars of a binary system is usually attributed to one of the following scenarios:

\begin{itemize}
\item Measurement of  different phases of long-term variations in both stars \citep[e.g.][]{1995ApJ...438..269B,1998ASPC..154..153B,1998ASPC..154.1235D}. This would correspond, for example, to having one star near a maximum and its companion near a minimum, being their average activity levels similar to each other. However, in this case we analyse time variations of both stars along years, being always $\zeta^{2}$ Ret more inactive than $\zeta^{1}$ Ret.
\end{itemize}

\begin{itemize}
\item Different rotational modulation of both stars \citep[e.g.][]{1998ASPC..154.1235D}. In order to avoid  the scatter probably associated to rotational modulation, we have used HARPS monthly means data. Therefore, the measurement of time-series shows that the stellar rotation is not the cause of these differences.
 \end{itemize}

\begin{itemize}

\item Differences in rotation rate \citep{1984ApJ...279..763N,2007ApJ...669.1167B,2008ApJ...687.1264M,2011ApJ...743...48W}.   
To analyse this possibility, we estimated both ``projected'' and  ``empirical'' rotation periods of $\zeta$ Ret, in order to compare them. The ``projected'' rotation periods were estimated following the formula 
P$_\mathrm{rot}$/$\sin$ $i$=(2$\pi$R$_{\star}$)/($v$ $\sin$ $i$), where $\sin$ $i$ is the inclination between the rotational axis and a perpendicular plane to the observer, R$_{\star}$ is the stellar radius and $v$ $\sin$ $i$ is the projected rotational velocity measured from the spectra. We note that P$_\mathrm{rot}$/$\sin$ $i$ is a lower limit to the rotational period. We used spectroscopic $v$ $\sin$ $i$ values of \citet{2003A&A...398..647R} and estimated R$_{\star}$ values from the Bayesian method through the PARAM code (0.87 $\pm$ 0.01 R$_{\odot}$ and 0.95 $\pm$ 0.01 R$_{\odot}$ for $\zeta^{1}$ Ret and $\zeta^{2}$ Ret, respectively). In this way, we obtained projected rotation periods of 16.4 $\pm$ 3.2 days and 17.9 $\pm$ 3.5 days for both stars, being similar within the errors. These uncertainties include the dispersion in the estimation of R$_{\star}$ and the errors associated to $v$ $\sin$ $i$ values.

Then, we derived the ``empirical'' rotation periods, i.e., rotation periods expected from MH08 calibration (see their equation 5) given the known activity and (B$-$V) colour. As a result, we obtained 13.2 $\pm$ 2.8 days and 16.5 $\pm$ 1.8 days for $\zeta^{1}$ Ret and $\zeta^{2}$ Ret, respectively. The errors include the uncertainties of log($\mathrm{R}'_\mathrm{HK}$), (B$-$V) and the scatter of the MH08 calibration. The empirical rotation period of $\zeta^{1}$ Ret seems to be shorter than the corresponding value to $\zeta^{2}$ Ret. However, when we consider the uncertainties they are indistinguishable.

When we compare both projected and empirical rotation periods, they are similar within the errors. Then, a rotation difference between $\zeta^{1}$ Ret and $\zeta^{2}$ Ret cannot be ruled out, due to their relatively large uncertainties. The uncertainties of the $v$ $\sin$ $i$ measurements are too large to conclusively show a difference in the rotation period.
We point out that both estimations show that $\zeta^{1}$ Ret rotates faster than $\zeta^{2}$ Ret. It would be desirable, for instance, to count with a suitable set of spectroscopic or photometric observations in order to better estimate a period by using the rotational modulation in a future work.

\end{itemize}

\begin{itemize}
\item MH08 point out that pronounced activity discrepancies ($\sim$0.1-0.2 dex) in coeval stars, could be associated to differences in (B$-$V) colour. Then, we wonder if it is possible to explain the observed activity difference between both components of $\zeta$ Ret. With this aim in mind, we compared this difference with those from the sample of near-identical pairs of MH08. In Fig.~\ref{nar00} we plotted $\Delta\log(R'_{HK})$ vs. $\Delta$(B$-$V), where the binary system $\zeta$ Ret is showed with a full circle. Following MH08, the order in the $\Delta$ differences correspond to $B-A$, being the B star those with the higher (B$-$V) value in the pair. We note that $\zeta$ Ret appears above all of the near-identical pairs of MH08. This would be a first indication of a somewhat rare behaviour of this pair. Then, we applied the MH08 activity-age calibration by adopting a common age of 4.5 $\pm$ 2.0 Gyr and a (B$-$V) difference of 0.04 $\pm$ 0.02, and compare these ``expected'' activity values with the observed ones. If we assume the greater possible age for the pair (i.e. 6.5 Gyr) and also the maximum possible (B$-$V) difference between them (0.06 mag), then we reach a maximum difference of 0.17 $\pm$ 0.05 dex in log($\mathrm{R}'_\mathrm{HK}$) (-5.11 and -5.28 dex for both stars). As a result, we find that the MH08 calibration would roughly reach the observed difference in log($\mathrm{R}'_\mathrm{HK}$) (0.220 $\pm$ 0.027 dex), but simultaneously estimating very low activity levels for each star, which is not observed. This shows that the $\zeta$ Ret binary system seems to be outside of the ``normal'' behaviour of the
near-identical pairs of MH08.

\begin{figure}
\centering
\includegraphics[width=\columnwidth]{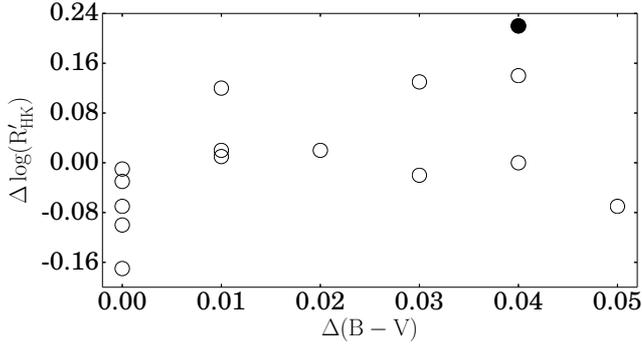}
\caption{Position of near-identical pairs of MH08 in the $\Delta$log($\mathrm{R}'_\mathrm{HK}$)$-\Delta$(B$-$V) diagram (empty circles). The position of the $\zeta$ Ret binary system, with the activity levels calculated in this work, is
also included in the figure (full circle).}
\label{nar00}
\end{figure}

\begin{figure}
\centering
\includegraphics[width=\columnwidth]{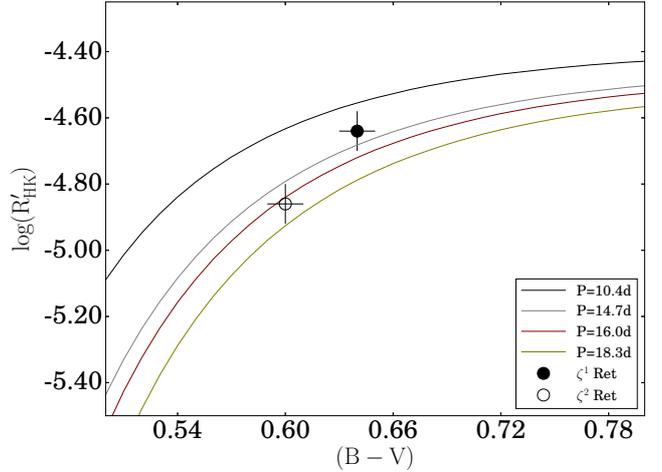}
\caption{Activity vs (B$-$V) for different rotational periods. Adopting P$_{1}$ and P$_{2}$ as the rotational periods of 
$\zeta^{1}$ Ret and $\zeta^{2}$ Ret, the colour curves correspond to periods of P$_{1}$ -1 $\sigma$ (10.4 days), P$_{1}$ + 1 $\sigma$ (14.7 days), P$_{2}$ -1 $\sigma$ (16.0 days) and P$_{2}$ + 1 $\sigma$ (18.3 days). We also indicate the position of $\zeta^{1}$ Ret (full circle) and $\zeta^{2}$ Ret (empty circle).}
\label{nar00fb}
\end{figure}

Also, in the Figure \ref{nar00fb} we plotted activity vs (B$-$V). Full and empty circles show the position of the stars $\zeta^{1}$ Ret and $\zeta^{2}$ Ret, respectively. Colour curves correspond to different rotation periods, according to MH08 calibration. 
Curves are shown for the $\pm$ 1 $\sigma$ region around the MHO8 empirical rotation period for $\zeta^{1}$ Ret and $\zeta^{2}$ Ret using the observed B-V and activity. When including the rotation periods uncertainties, we obtain a region rather than a single curve for each star. 

Previously, we mentioned that the MH08 calibration suggests rotation periods of 13.2 $\pm$ 2.8 and 16.5 $\pm$ 1.8 days, compatible with the activity levels of the stars  $\zeta^{1}$ Ret and  $\zeta^{2}$ Ret, respectively. If we suppose that both stars do present the same rotation period (e.g. $\sim$15 days), this would imply mean activity levels of -4.69 $\pm$ 0.06 and -4.80 $\pm$ 0.06 dex for both stars i.e. a difference of $\sim$0.11 $\pm$ 0.12 dex. Then, it seems unlikely to explain the measured activity difference of 0.220 dex only by adopting a difference in (B$-$V) between both stars. For instance, as can be seen in the Figure \ref{nar00fb}, the stars $\zeta^{1}$ Ret and $\zeta^{2}$ Ret are located above and below of the curves corresponding to periods of 16.0 days and 14.7 days. This would point toward a different rotational periods for both stars. Then, the MH08 calibration could (in principle) explain the difference observed in activity, by considering the compound effect of (B$-$V) together with the assumption of different rotational periods for the stars (the mentioned 13.2 $\pm$ 2.8 and 16.5 $\pm$ 1.8 days, respectively). However, to our knowledge there is no observational evidence of different rotational rates in these stars. A better estimation of the rotational periods would help to support (or rule out) this compound scenario trying to explain the activity difference in this binary system.

In the light of the results obtained in this work, it is clear that the case of the $\zeta$ Ret binary system
is a challenge for activity-age calibrations: the greater contrast in their chromospheric activity levels, the more difference in their stellar ages. Following the calibration of MH08, the difference in chromospheric age between both stars amounts
to $\sim$2 Gyr, which clearly exceeds the expected dispersion of $\lesssim$1 Gyr for the case of binary stars \citep{1998ASPC..154.1235D,2008ApJ...687.1264M}. The $\zeta$ Ret binary system is considered by MH08 in the group of ``near identical'' pairs (see their Table 3). We mentioned above that the difference in (B$-$V) colour alone could not explain the activity levels  of both stars. Thus, when dealing with stellar clusters, which are usually used for activity-age studies, one should carefully check not only for (B$-$V) colour, mass, and metallicity biases, but also for long-term high activity-differences such as those found in $\zeta$ Ret.

\end{itemize}

\begin{itemize}
\item Evolved stars have significantly lower chromospheric activity levels compared to main sequence stars \citep[see e.g.][for details]{2004AJ....128.1273W}. Then, the possibility that only $\zeta^{2}$ Ret evolve off the main sequence is other tentative cause for its lower activity level. The estimated surface gravity of this star is $\log g=$ 4.54 \citep{2016A&A...588A..81S}. \citet{2004AJ....128.1273W} defines as ``unambiguously evolved'' those stars with $\Delta$M$_\mathrm{v} >$ 1, where $\Delta$M$_\mathrm{v} $ is the difference between the absolute magnitude M$_\mathrm{v}$ and the function M$_\mathrm{v}$ (B$-$V), which is a fit to main-sequence stars as a function of the colour (B$-$V). Similar to the plot of \citet{2004AJ....128.1273W}, in the Figure \ref{nar06} we present M$_\mathrm{v}$ vs (B$-$V) for the stars included in the work of \citet{2005ApJ...622.1102F}. The triangles correspond to main-sequence stars (fitted using a continuous red line), while the squares correspond to stars that satisfy the Wright's criteria as evolved. The circles (black and white) show the position of the stars $\zeta^{1}$ Ret and $\zeta^{2}$ Ret with $\Delta$M$_\mathrm{v}$ values of $\sim$ -0.40 and $\sim$ -0.37 mag, respectively. Then, there is no evidence for an evolved status for the stars in this binary system.

\begin{figure}
\centering
\includegraphics[width=\columnwidth]{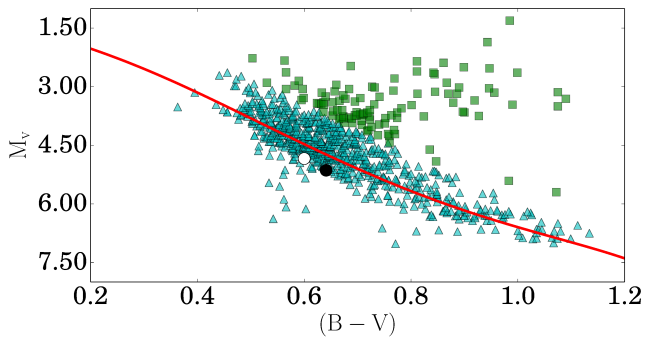}
\caption{Position of $\zeta^{1}$ Ret (black circle) and $\zeta^{2}$ Ret (white circle) stars in the M$_\mathrm{v}$ $-$ (B$-$V)
diagram. The continuous red line shows the fit to main-sequence stars (triangles). Evolved stars are indicated with squares.}
\label{nar06}
\end{figure}

\end{itemize}

\begin{itemize}
\item The separation of 3700 AU between both stars \citep{2001AJ....122.3466M} allow us to discard a possible interaction between
$\zeta^{1}$ Ret and $\zeta^{2}$ Ret. The non-detection of planets orbiting both stars \citep[e.g.][]{2008A&A...487..373S,2013A&A...552A..78Z} rule-out a possible SPI (star-planet interaction) activity effect.

It is intriguing the presence of a dust disk orbiting around $\zeta^{2}$ Ret \citep{2008ApJ...674.1086T,2010A&A...518L.131E}. A possible (current) interaction between the disk and the star seems unlikely, due to the separation between them \citep[$\sim$100 AU, ][]{2010A&A...518L.131E}. On the other hand, it is difficult to determine if the presence of a dust disk could possibly alter the rotational evolution of the stars. To our knowledge, there is no statistical study comparing the rotational rates with the presence of a dust disk.

\end{itemize}

\begin{itemize}
\item \citet{1980PASP...92..385V} noted that solar neighbourhood stars could be roughly divided into two populations,
namely active and inactive stars. The separation between these populations is made by the so-called Vaughan-Preston gap
(hereafter VP gap), located around log($\mathrm{R}'_\mathrm{HK}$) $\sim-4.75$ dex, which is a region of intermediate activity
containing very few stars \citep{1980PASP...92..385V,1996AJ....111..439H,2009A&A...499L...9P}.
\citet{1996AJ....111..439H} suggest that this region is a transition zone rather than a gap. The stars $\zeta^{1}$ Ret and $\zeta^{2}$ Ret present log($\mathrm{R}'_\mathrm{HK}$) values of $-4.64$ dex and $-4.86$ dex, respectively,
being then considered as active and inactive stars on different sides of the VP gap.

Different literature works tried to explain the presence of the VP gap.
For instance, \citet{1998MNRAS.298..332R} suggest an abrupt change in metallicity between both populations of active and
inactive stars, however the components of our binary system present almost identical metal content \citep{2016A&A...588A..81S}.
\citet{2002A&A...394..505B} suggest that the VP gap represents a transition from a multiple-mode
dynamo to a single-mode dynamo. Then, \citet{2004A&A...426.1021P} proposed that most active stars are usually young,
and then a fast decay of chromospheric activity occurs roughly between 0.6 and 1.5 Gyr of main sequence lifetime,
after which a kind of plateau appears. They propose that the abrupt decline in activity with age could
explain the VP gap. Then, \citet{2009A&A...499L...9P} propose that the chromospheric activity does not evolve smoothly
with time: stars change from active to inactive crossing the VP gap on time-scales that might be as short as 200 Myr.

Following this idea, we can interpret that only $\zeta^{2}$ Ret would have very recently crossed the VP gap
(in the last $\sim$200 Myr) while $\zeta^{1}$ Ret is close to cross it. We consider that this is a plausible scenario,
which has never been proposed before to explain a difference of activity in a binary system. If this is the case, 
both stars should present an age lower than $\sim$1.5 Gyr, which corresponds to the time when most stars seem to cross
the VP gap, according to \citet{2009A&A...499L...9P}.
However, the estimated common age for $\zeta$ Ret is $\sim$4.6 yr, according to the isochrone method. Then, if additional binary stars straddling the gap were found with ages $>$ 1.5 Gyr (as $\zeta$ Ret binary system appears to be), it would be a reason to be suspicious of the Pace \& Pasquini's result that stars cross the gap at ages between 0.6 and 1.5 Gyr.

\end{itemize}

\begin{itemize}
\item

It has been suggested that the existence of a remarkable difference in the activity behaviour among binary components, could be associated to a similar MM state in the star with lower activity \citep{1998ASPC..154.1235D,2004AJ....128.1273W}. However, current criteria to identify MM candidates are still under discussion \citep[e.g.][]{2004AJ....128.1273W,2007ApJ...663..643J}.

The first stellar analogies to the MM come from stars with relatively constant activity levels, called ``flat'' stars ($\sigma_{\mathrm{S}}$/$\overline{\mathrm{S}}$ $<$ 1.5\%) in \citet{1990Natur.348..520B} and \citet{1995ApJ...438..269B}. 
It is interesting to note that $\zeta^{2}$ Ret satisfies the Baliunas's criteria for a flat star ($\sigma_{\mathrm{S}}$/$\overline{\mathrm{S}}\sim1.4\%$) and for a cycling star simultaneously (FAP $\leq 10^{-2}$). Then, some authors 
considered as MM candidates those stars with log($\mathrm{R}'_\mathrm{HK}$)  $<$ $-5.1$ dex \citep[e.g.][]{1996AJ....111..439H}.
Due to this criterion was established from a sample where almost all were evolved stars \citep{2004AJ....128.1273W}, it has been proposed that this low activity value should be higher than $-5.1$ dex, and that UV and X-ray data could also be used to identify MM candidates \citep[][]{2007ApJ...663..643J}. In addition, 
\citet{2004AJ....128.1273W} suggested that MM states in main-sequence stars should not be constrained to the log($\mathrm{R}'_\mathrm{HK}$) $<$ $-5.1$ dex condition, proposing that (section 4.1) a higher value of log($\mathrm{R}'_\mathrm{HK}$) $\sim$ $-5.0$ dex may represent the minimum level of activity in main sequence stars. Then, the author suggest the search for absence of activity variations or appreciable activity differences between components of binary systems as an useful test to identify MM states. Considering that $\zeta^{1}$ Ret and $\zeta^{2}$ Ret satisfy this final condition, we explored the possibility that the low activity of $\zeta^{2}$ Ret could be attributed to a similar solar MM phase.

It is well known that during this period, the sunspot number\footnote{The sunspot number, a key indicator of solar activity, has a strong correlation with the \ion{Ca}{ii} H\&K chromospheric activity proxy \citep[e.g.][]{2016SoPh..291.2967B}.} was extremely reduced, although they did not disappear 
\citep{1993A&A...276..549R}. 
In addition, some evidence has been found suggesting that solar cycle was still in progress during the MM 
\citep[e.g.][]{1998SoPh..181..237B,2003mmvs.book.....S,2015A&A...577A..71V}. After the solar MM end, it was followed by a
gradual increase in cycle amplitudes of the cyclic variability \citep{2015LRSP...12....4H}. In this way, following the interpretation of \citet{1998ASPC..154.1235D} and \citet{2004AJ....128.1273W}, maybe  
$\zeta^{2}$ Ret is not in a MM state, instead, this component could be emerging from (or going to) it and at the same time, showing a stellar activity cycle with a period similar to the solar case.

We tried to compare the average difference in activity between $\zeta^{1}$ Ret and $\zeta^{2}$ Ret ($\sim$0.220 $\pm$ 0.027 dex) with an estimation of the decrease in activity of the Sun during its MM state. This is a difficult task, given that there are
only estimations and no direct measurements of solar chromospheric index in the mentioned period.
\citet{2017ApJ...835...25E} found a mean activity of the Sun of log($\mathrm{R}'_\mathrm{HK}$) $=$ $-4.9427 \pm 0.0072$ by taking data from
the cycles 15 to 24. If we adopt for the Sun the value of log($\mathrm{R}'_\mathrm{HK}$) $\sim$ $-5.0$ suggested by \citet{2004AJ....128.1273W} as the 
minimum level of activity in main sequence stars, then we roughly estimate a decrease of $\sim0.06$ dex for the Sun. 
This would indicate that the activity difference between $\zeta^{1}$ Ret and $\zeta^{2}$ Ret would be possibly greater than the activity decrease of the Sun in its MM state. Then, a possible decrease in activity in the star $\zeta^{2}$ Ret could not be ruled out.

In order to search for evidence of this behaviour from an independent source, we also explored the available X-ray data\footnote{As previously indicated, these individual observations were taken at different dates} following the \citet[][]{2007ApJ...663..643J} suggestion. To do so, we estimated F$_X$ values of (5.11 $\pm$ 0.08) 10$^{-13}$  and (0.25 $\pm$ 0.32) 10$^{-13}$ erg s$^{-1}$ cm$^{-2}$ for $\zeta^{1}$ Ret and $\zeta^{2}$ Ret in the 2-10 KeV band. This seems to indicates that $\zeta^{1}$ Ret is more intense than $\zeta^{2}$ Ret, as also suggested by the \ion{Ca}{ii} H\&K fluxes. Then, the difference between the stellar activity for both components and the activity cycle detected here, 
allow us to suppose that $\zeta^{2}$ Ret is possibly emerging from a similar MM state. In this way, before the stellar cycle that we are now observing ($\sim$10 yr), perhaps the mean stellar activity of $\zeta^{2}$ Ret was even lower. We stress, however, that long-term \ion{Ca}{ii} H\&K and X-ray data (including for the Sun) would be desirable in order to more properly identify $\zeta^{2}$ Ret as a MM candidate.

\end{itemize}

\section{Conclusions} 
\label{sec.concl}
We performed, for the first time, a detailed long-term activity study of the binary system $\zeta$ Ret
by using HARPS spectra, covering measurements between the years 2003 and 2016.
Both stars are physically very similar between them and also both are solar analogues.
We detected a periodic modulation of  $\sim$10 yr in the star $\zeta^{2}$ Ret with a mean
$S$ index (0.180) similar to that of the Sun. We also note that this object satisfies previous criteria for a flat star and for a cycling star simultaneously. On the other hand, $\zeta^{1}$ Ret showed a higher activity level
(in \ion{Ca}{ii} H\&K and X-ray) with an erratic rather than a cyclic modulation.

We discussed possible scenarios in order to account for the activity difference
observed in the $\zeta$ Ret binary system. On one side, a different rotational rate between both stars cannot be totally
ruled out as a possible cause. However, this would require a more precise determination
the rotation periods of $\zeta^{1}$ Ret and $\zeta^{2}$ Ret. On the other hand, we showed that the difference in (B$-$V) alone is unlikely to explain the activity difference, in the context of the activity-colour-rotation MH08 calibrations.
However, adopting the compound effect of a difference in (B$-$V) of 0.06 mag (at the limit of the observed 0.04 $\pm$ 0.02 mag), together with the assumption of different rotational periods (estimated as 13.2 $\pm$ 2.8 and 16.5 $\pm$ 1.8 days for both stars) could in principle explain the activity difference observed. Finally, we also propose that the star $\zeta^{2}$ Ret is possibly emerging from (or going to) an state similar to the MM, as another possible scenario. We wonder if a similar activity difference could be also observed in other binary systems, which would help to verify or rule out these possible scenarios. This require to continue monitoring this and other binary systems, if they are available, with new HARPS data. In addition, these measurements can be complemented with observations from the Complejo Astron\'omico El Leoncito (CASLEO) observatory, also situated in  the southern hemisphere.

Finally, when dealing with activity-age calibrators, we suggest to carefully check not only for B$-$V colour, mass and metallicity, but also for possible long-term activity variations such as those found in the $\zeta$ Ret binary system.

\section*{Acknowledgements} 
M. Flores and M. Jaque Arancibia acknowledge the financial support from CONICET in the forms of Post-Doctoral Fellowships. We also 
cordially thank the referee Dr. Ricky Egeland for a careful reading and constructive comments that greatly improved the work.

\bibliographystyle{mnras}
\bibliography{references}



\clearpage
\begin{appendix}

\end{appendix}

\end{document}